\documentclass[aps,pre,twocolumn,superscriptaddress,floatfix,10pt]{revtex4}
\usepackage{graphicx}
\usepackage{amsmath}
\usepackage{color}
\usepackage{wrapfig}
\usepackage{latexsym}
\begin{document}

\title{Tuning thermal transport in highly cross-linked polymers by bond induced void engineering}% networks}

\author{Debashish Mukherji}
\email[]{debashish.mukherji@ubc.ca}
\affiliation{Quantum Matter Institute, University of British Columbia, Vancouver BC V6T 1Z4, Canada}
\author{Manjesh Kumar Singh}
\affiliation{Department of Mechanical Engineering, Indian Institute of Technology Kanpur, Kanpur UP 208016 India}

%\date{\today}

\begin{abstract}
Tuning the heat flow is fundamentally important for the design of advanced 
functional materials. Here, polymers are of particular importance because
they provide different pathways for the energy transfer. More specifically,
the heat flow between two covalently bonded monomers is over 100 times faster than between
the two non-bonded monomers interacting via van der Waals (vdW) forces.
Therefore, the delicate balance between these two contributions often provide a guiding 
tool for the tunability in thermal transport coefficient $\kappa$ of the polymeric materials.
Traditionally most studies have investigated $\kappa$ in the linear polymeric materials, 
the recent interests have also been directed towards the highly cross-linked polymers (HCP). 
In this work, using the generic molecular dynamics simulations we investigate the factors effecting $\kappa$ of HCP. 
We emphasize on the importance of the cross-linking bond types and its influence on the network microstructure
with a goal to provide a guiding principle for the tunability in $\kappa$. 
While these simulation results are discussed in the context of the available experimental data, we also make predictions.	
\end{abstract}

\maketitle

\section{Introduction}

Polymers are an important class of high entropy materials that are extremely important for our everyday life \cite{cohen10nm,mukherji20arcmp}, 
finding possible applications ranging from the household items \cite{henry14rev,mehra18amt}, electronic packaging \cite{pipe14nm},
organic solar cells and light emitting diodes \cite{kim05apl,cola16aami}, thermoelectrics \cite{wang17afm,liu19mme,park20apmi}, 
and for the defense materials \cite{pal13apolsc,sirk16sm}. The standard architecture of a polymer chain consists of a string of covalently bonded monomers
that exhibit interesting properties at different length, time and energy scales \cite{cohen10nm,mukherji20arcmp}.
Another class is when each monomer can form multiple bonds with its neighbors,
commonly known as the highly cross-linked polymers (HCP). Because of the network connectivity, 
HCPs are light weight high performance materials that often exhibit extraordinary and unexpected mechanical behavior \cite{pal13apolsc,sirk16sm,stevens01mac,mukherji08pre}. 
Here, one quantity that is intimately linked to the mechanical response of a materials is their thermal transport coefficient $\kappa$ \cite{cahill92prb,braun18am}.
In this context, understanding the heat propagation in bulk polymers is exceedingly challenging because of 
their complex microstructure, while has tremendous potential in designing advanced functional materials with tunable properties \cite{pipe14nm,xie16mac,mukherji19prm}.

Studying the heat flow in polymers is microscopically interesting because at the monomer level there are two distinct pathways
for the energy transfer, i.e., between two covalently bonded monomers and between non-bonded neighbors dictated by the 
van der Waals (vdW) contacts \cite{chen08prl,shen10natnano,liu12prb,luo13acsnano}. The strength of vdW interaction is less than $k_{\rm B}T$, while 
it is $80k_{\rm B}T$ for the covalent bonds \cite{mukherji20arcmp,deju}. Here, $k_{\rm B}$ is the Boltzmann constant and $T = 300$ K. 
Owing to this stark contrast in the relative interaction strengths, the material stiffness also changes 
from about $5$ GPa for the vdW interactions \cite{xie16mac} to even higher than $250$ GPa for a covalent bond \cite{crist96} and thus leads to 
a contrast $\kappa_{\rm covalent}/\kappa_{\rm vdW} > 100$ \cite{shen10natnano}. 
In the case of the hydrogen bonding between the non-bonded monomers, the strength of which is about $4k_{\rm B}T$, 
$\kappa_{\rm H-Bond}/\kappa_{\rm vdW}\simeq 2-4$ \cite{xie16mac,mukherji19prm} and thus also 
reduces $\kappa_{\rm covalent}/\kappa_{\rm H-bond}$. Therefore, the higher $\kappa$ between 
two bonded neighbors will automatically infer that $\kappa$ for a bulk polymeric material can be increased by 
increasing the bond density $\rho_{\rm b}$, i.e., a higher value of $\rho_{\rm b}$ will be expected to significantly 
increase $\kappa$. Here, a prototypical system is the HCP where the physical properties are dominated by the three$-$dimensional 
network of bonds \cite{pal13apolsc,sirk16sm,stevens01mac,mukherji08pre}. Moreover, the experiments have shown that this standard 
understanding does not always hold, instead $\kappa$ with different cross-linking types exhibits rather anomalous behavior \cite{vars09pol,cahill20acs,huo20pccp}. 
For example, most HCP can only show an improvement in $\kappa$ by about $1.1-1.3$ times in comparison to the standard linear 
polymeric materials, while $\kappa$ is even lower for some HCPs than the linear chains \cite{cahill20acs}.

Traditionally, most simulation studies on HCP are devoted to investigate the network structures and their mechanical properties. 
These include from the generic \cite{stevens01mac,mukherji08pre} to multiscale \cite{ara16jpcb,per20mac}, and to all-atom \cite{sirk16sm,khalatur07mac} 
molecular dynamics simulations. Recent interests are also devoted to study the thermal transport of HCP \cite{vars09pol,huo20pccp}.
In this context, while the all-atom simulations are useful for the quantitative comparisons with the experimental data, 
they also pose significant challenge from the computational perspective. For example, due to the lack of the microstructural network details from 
the realistic systems$-$ curing is always nontrivial, restricting the
accessibility to a wide range of (macro-)molecular structures, and often are limited the goodness of the 
force field parameters. Therefore, the generic model can provide a better alternative 
to give the underlying physical mechanism and also provides the necessary flexibility 
for the parameter tuning in these complex systems \cite{stevens01mac,mukherji08pre,huo20pccp}. 
Motivated by these observations, the goal of this work is to investigate the effects of bonding on the 
network microstructure of HCP and $\kappa$ with an aim to propose a generic scheme that can serve as a guiding principle 
for the future experimental studies.

The remainder of the paper is organized as follows: In Section \ref{sec:method}, we sketch our methodology. 
Results and discussions are presented in Section \ref{sec:result}, and finally, the conclusions are
drawn in Section \ref{sec:conc}.

\section{Medel and method}
\label{sec:method}

For this study we have chosen a set of neat HCP with different network functionality $n$, i.e., each monomer can form a 
maximum of $n$ bonds. Here we choose linear polymer melt (i.e., $n=1$) with a chain length $N_{\ell} = 50$, 
tri-functional HCP (i.e., $n = 3$), and tetra-functional HCP (i.e., $n = 4$). In all these systems, the
total number of monomers in a simulation box is taken as $N = 2.56 \times 10^5$.

\subsection{Interaction potentials}

We employ a generic molecular dynamics simulation approach. Here, all non-bonded interactions are modelled using a
6$-$12 Lennard-Jones (LJ) potential,
\begin{equation}
	u_{\rm non-bonded} = 4\epsilon \left[\left(\frac {\sigma}{r}\right)^{12} - \left(\frac {\sigma}{r}\right)^6 \right]~ {\rm for}~ r < 2.5\sigma.
\end{equation}
Here, $\epsilon$ and $\sigma$ are the LJ energy and LJ length, respectively. 
This leads to a unit of time $\tau = \sigma \sqrt{m/\epsilon}$, with m being the mass of the monomers. 
Our systems consist of $N = 2.56 \times 10^5$ LJ particles randomly distributed within a cubic box 
at an initial monomer number density $\rho_{\rm m} = 0.85\sigma^{-3}$. 
The equations of motion are integrated using the velocity Verlet algorithm with a time step $0.005\tau$ and the 
temperature is set to $T = 1\epsilon/k_{\rm B}$, thus representing a gel phase. The temperature is imposed using a Langevin thermostat with a 
damping coefficient of $\gamma = 1\tau^{-1}$. The initial LJ system is equilibrated for $10^6$ steps.
For the chain connectivity, we have used two different bond types, 
namely the finitely extensible nonlinear elastic (FENE) and the harmonic potentials.

\subsubsection{FENE bond}
\label{fene}

A bond between two monomers is defined by the combination of repulsive 6-12 LJ potential,
\begin{equation}
u_{\rm bonded} = 4\epsilon_{\rm b} \left[\left(\frac {\sigma_{\rm b}}{r}\right)^{12} - \left(\frac {\sigma_{\rm b}}{r}\right)^6 
	+ \frac {1}{4}\right]~ {\rm for}~ r < 2^{1/6}\sigma_{\rm b},
\end{equation}
and the FENE potential \cite{kgmodel},
\begin{equation}
	u_{\rm FENE} = -\frac {1} {2} k R_{\circ}^2 \ln \left[1 - \left(\frac {r}{R_{\circ}}\right)^2\right].
\end{equation}
Here, $\epsilon_{\rm b}$ and $\sigma_{\rm b}$ are the LJ interaction energy and the LJ interaction length between bonded monomer, respectively. 
While the default FENE bond parameters, $k = 30\epsilon/\sigma^2$ and $R_{\circ} = 1.5\sigma$, give a typical bond 
length of $\ell_{\rm b} \simeq 0.97\sigma$ \cite{kgmodel}, we have also parameterized the FENE interaction for different $\ell_{\rm b}$. 
The details of parameters are listed in the Supplementary material \cite{epaps}.
These parameters ensured that the FENE bond stiffness remains reasonably invariant with the changing $\ell_{\rm b}$.

\subsubsection{Harmonic Bond}
\label{harmonic}

Most commodity polymers, such as polystyrene (PS), polyethylene (PE), poly(methyl methacrylate) (PMMA), poly(N-isopropyl acrylaime) (PNIPAm), 
poly(acrylic acid) (PAA), and poly(acrylamide) (PAM), the backbone connectivity is dictated by the carbon-carbon 
covalent bond \cite{mukherji19prm} and in polycarbonate these are benzene rings \cite{bi1,bi2}. Both these bonds are significantly stiffer than the 
most cross-linkers used to synthesize HCP \cite{cahill20acs,palmese14jmat}. Therefore, to mimic the stiff bonds, 
we have used harmonic potential,
\begin{equation}
u_{\rm harmonic} = \frac {\epsilon}{2s^2} \left(r - \ell_{\rm b}\right)^2,
\end{equation}
with $s$ being the standard deviation of the bond length fluctuation \cite{bi1,bi2}.
%For this purpose, we have taken $s = 0.0126\sigma$ representing the linkage in polycarbonate \cite{abrams01jcp} and for $\ell_{\rm b}=0.90\sigma$, $0.97\sigma$,
%$1.05\sigma$, and $1.10\sigma$. 
%As noted earlier \cite{abrams01jcp}, we also wish to highlight that the fluctuation of 
%the FENE bond is about 3.4\% at $T=1\epsilon/k_{\rm B}$, while this is only about 1\% for the harmonic bond with $s = 0.0126\sigma$.

\subsection{Thermal transport calculations}

The thermal transport coefficient $\kappa$ is calculated using the Kubo-Green
method \cite{greenkubo} implementation in LAMMPS \cite{lammps}. The equations of motion are integrated in the microcanonical ensemble.
The heat flux autocorrelation function
%\begin{equation}
$C(t) = \langle {\bf J}(t) \cdot {\bf J}(0) \rangle$
%\end{equation}
is obtained by sampling the heat flux vector ${\bf J}(t)$. 
The typical $C(t)$ data is shown in the Supplementary Fig. S1 \cite{epaps}.
Here we choose a sampling period of $0.005{\tau}$ to determine the correlation function over 
a time frame of $0\leq\ t \leq 50 {\tau}$, which is one order of magnitude larger than the 
typical de-correlation time, see the Supplementary Fig. S1 \cite{epaps}.
During a total simulation of $5 \times 10^4 {\tau}$, we accumulate correlation data and compute a running average of ${C}(t)$
Finally, $\kappa$ values are calculated by taking the plateau of the Green-Kubo integral for the component along the chain,
\begin{equation}
	\kappa=\frac{v}{3k_{{\rm B}}T^{2}}\int_0^{\mathcal{T}} C(t) {\rm d}{t},
	\label{eq:kgint}
\end{equation}
where $v$ is the system volume. Ideally the sampling time $\mathcal{T} \to \infty$. 
Here, however, we calculate $\kappa$ by taking an average between $30\tau \le \mathcal{T} \le 50\tau$ from 
the plateau of the cumulative integral in Eq. \ref{eq:kgint}.

\section{Results and discussions}
\label{sec:result}

\subsection{Sample cure}
\label{sec:netcur}

The bonds are formed during a network curing stage using a similar protocol used earlier \cite{mukherji08pre}. 
Within this protocol, starting from a LJ melt at $T = 1\epsilon/k_{\rm B}$,
bonds are randomly formed between a pair of particles when: 1) two particles are closer than $1.1\ell_{\rm b}$ distance, 
2) they have not formed the maximum number of allowed bonds $n$, and 3) a random number between 
zero and one is less than the bond forming probability of 0.05.  
Unless stated otherwise, the network curing is performed for $10^6$ time steps (or equivalent of $t_{\rm cure}=5\times 10^3\tau$) 
under the canonical simulation. 

\begin{figure}[ptb]
\includegraphics[width=0.49\textwidth,angle=0]{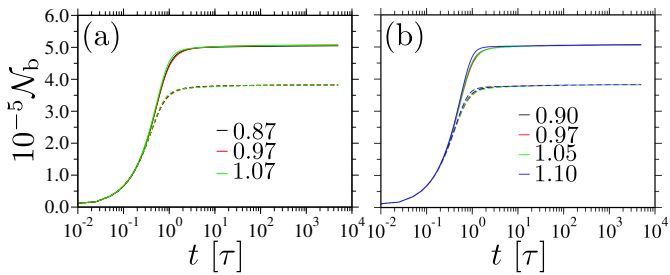}
	\caption{The formation of the total number of bonds $\mathcal{N}_{\rm b}$ with time $t$ for different bond lengths $\ell_{\rm b}$. 
	The data is shown for $n=3$ tri-functional (dashed lines) and $n=4$ tetra-functional (solid lines) monomers. 
	Parts (a) and (b) show the data for the FENE and harmonic bonds, respectively.
	\label{fig:bondform}}
\end{figure}

\begin{figure}[ptb]
\includegraphics[width=0.43\textwidth,angle=0]{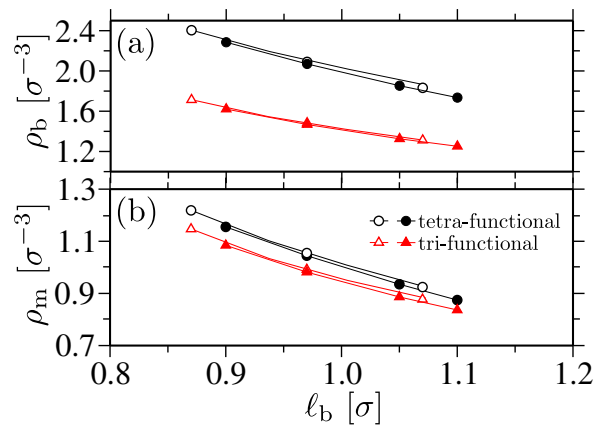}
	\caption{Number density of bonds $\rho_{\rm b}$ (panel a) and monomers $\rho_{\rm m}$ (panel b) as a
	function of bond length $\ell_{\rm b}$. The data is shown for the systems after the isobaric equilibration.
	Open and solid symbols are for the FENE and the harmonic bonds, respectively.
        Lines are drawn to guide the eye.
\label{fig:dens}}
\end{figure}

Fig. \ref{fig:bondform} shows the formation of the total number of bonds $\mathcal{N}_{\rm b}$ during the 
curing stage for different $\ell_{\rm b}$ and $n$ for both bond types. It can be seen that $\mathcal{N}_{\rm b}$ values reach a plateau maximum around $t \simeq 10^2\tau$.
We have also calculated the percentage of cure $\mathcal{C}$ after $t_{\rm cure}$, see the Supplementary Fig. S2 \cite{epaps}. 
It can be appreciated that all systems reach over 98.5\% cure. 
Furthermore, $\mathcal{C}$ increases with $\ell_{\rm b}$ by a factor of less than 1\%.
This slight increase in $\mathcal{C}$ is due to the pure geometric arrangements where a longer $\ell_{\rm b}$ leads to a 
larger number of nearest neighbors and thus on average forming a larger number of bonds.
Moreover, we find that the number densities of bonds $\rho_{\rm b}$ (see part a of Fig. \ref{fig:dens}) and 
monomers $\rho_{\rm m}$ (see part b of Fig. \ref{fig:dens}) decrease by 30-35\% with increasing $\ell_{\rm b}$.
We will come back to this density effect at a later stage of this manuscript.

\begin{figure*}[ptb]
\includegraphics[width=0.73\textwidth,angle=0]{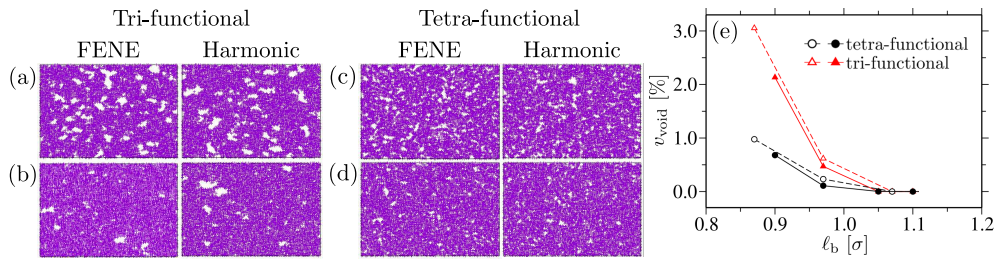}
	\caption{Panels (a-d) show the simulation snapshots of a $2\sigma$ thick layer along the $z$ direction 
	after the network cure in the canonical ensemble. The lateral dimensions of the snapshots are $67.03 \sigma$.
	Snapshots are shown for different bond lengths $\ell_{\rm b}$ 
	and functionalities $n$, as mentioned in the figure headings. Panels (a) and (c) are for $\ell_{\rm b} = 0.90\sigma$ and 
	panels (b) and (d) are for $\ell_{\rm b} = 0.97\sigma$. Panel (e) shows the fraction of the void volume $v_{\rm void}$
	as a function of $\ell_{\rm b}$. 
	Open and solid symbols are for the FENE and the harmonic bonds, respectively.
        Lines are drawn to guide the eye.
\label{fig:snapvoid}}
\end{figure*}

Another interesting feature of the microstructure of these networks are the formation of 
rather large voids immediately after the cure, see Figs. \ref{fig:snapvoid}(a-d). 
The void formation in these systems are not surprising given that two adjacent monomers can form all $n$ bonds within 
a small solid angle pointing away from each other, while these two native monomers may or may not form bonds within themselves \cite{mukherji08pre}. 
Furthermore, these void structures are dictated by the system thermodynamics$-$ where starting from a homogeneous 
monomeric mixture at a given $\rho_{\rm m}$, the formation of a void with a particular size is dictated by the competition 
between the entropy penalty and the surface energy reduction. 
In this context, it has been previously shown these network microstructural features can lead to the interesting mechanical 
response of the HCP networks \cite{sirk16sm,palmese14jmat}. 

The void sizes become larger with decreasing $\ell_{\rm b}$ 
(see the comparison between panels a and c in Fig. \ref{fig:snapvoid})
and $n$ (see the comparison between panels a and b in Fig. \ref{fig:snapvoid}). 
We have also calculated the void sizes in these systems
using a protocol proposed earlier \cite{mukherji08pre}. In this protocol,
a simulation domain is divided into cubic voxels with $1\sigma$ box lengths. 
A voxel is considered to be a void if a particle is not within a distance $0.5\sigma$ from the voxel boundary. 
For $\ell_{\rm b}=0.90\sigma$, we find that the 
largest void is about $10-15\sigma^3$ for $n=4$ and $\sim 60\sigma^3$ for $n=3$.
Moreover, for $\ell_{\rm b} \ge 1.05\sigma$ voids are not present that is predominantly because of the 
homogeneous bond formation above a critical bond length. We also wish to note that the fractions of the total void volume $v_{\rm void}$ 
are on average smaller for the harmonic bonds in comparison to the FENE bonds, see Fig. \ref{fig:snapvoid}(e).
This observation is also not surprising given that in our model, by construction, the harmonic bonds 
are stiffer than the FENE bonds.

After the curing procedure, the systems are subsequently equilibrated in the isobaric ensemble at zero pressure. 
During this process the voids can collapse forming several protovoids (void centers) within the sample \cite{mukherji08pre}.
The detailed structures are shown in the Supplementary Section S2 and Figs. S3-S6 \cite{epaps}.

To summarize the network cure procedure, we find three key features:
\begin{itemize}

\item Percentage of network cure $\mathcal{C}$ increases by $\simeq 1\%$ with the bond length $\ell_{\rm b}$.

\item Number density of bonds $\rho_{\rm b}$ and monomers $\rho_{\rm m}$ decreases by $\simeq 30 - 35\%$ with $\ell_{\rm b}$ giving 
	free volume and dilution of bonds.	

\item Fraction of the total void volume $v_{\rm void}$ decreases with $\ell_{\rm b}$, reaching a vanishing value 
	for $\ell_{\rm b}\ge 1.05\sigma$.

\item For $\ell_{\rm b}>1.05$ the bond formation in a network is rather homogeneous.

\end{itemize}
We will now show how the delicate balance between $\mathcal{C}$, $\rho_{\rm b}$, $v_{\rm void}$ and 
the bond stiffness can act as a guiding principle for $\kappa$ tunability in cross-linked networks. 

\subsection{Thermal conductivity}
\label{subsec:therm}

\begin{figure}[ptb]
\includegraphics[width=0.43\textwidth,angle=0]{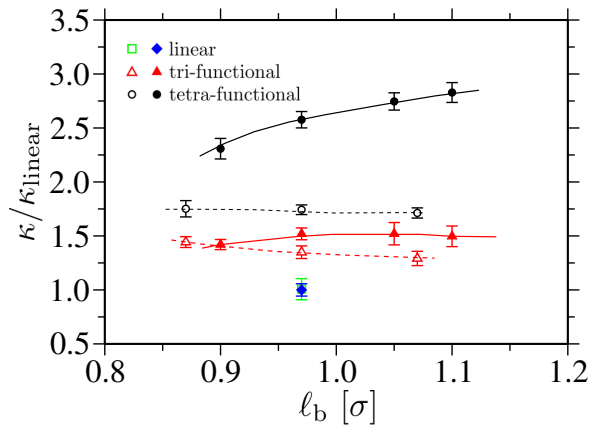}
	\caption{The normalized thermal transport coefficient $\kappa/\kappa_{\rm linear}$ as a function of the bond length $\ell_{\rm b}$.
	The data is shown for different network functionality $n$ and for the both bonds. The data is normalized with respect to the linear chain 
	connected by the harmonic springs giving $\kappa_{\rm linear} = 5.4\pm0.3~k_{\rm B}/\tau {\rm m}$.
	Here the linear chain lengths are chosen as $N_{\ell}=50$.
	Open and solid symbols are for the FENE and the harmonic bonds, respectively.
        Lines are drawn to guide the eye.
\label{fig:kappalb}}
\end{figure}

We will now discuss the most important results of this manuscript, namely the variation of $\kappa$ with 
different system parameters, see Fig. \ref{fig:kappalb}.

It can be seen for the linear chains that $\kappa$ remains invariant for both bonds, 
see the green $\Box$ and blue $\diamond$ data points in Fig. \ref{fig:kappalb}.
Note that the non-bonded interactions are identical in these two model melts.
Ideally, the stiffer bonds can lead to higher $\kappa$, i.e., the heat transfer between two bonded monomers 
can increase with bond stiffness. Moreover, the typical length scale over which this increased $\kappa$ can be observed 
(or the mean free path of the heat flow) is only about 2-3 monomer units (i.e., the typical segment length) \cite{mukherji19mac}. 
Over the full chain backbone, however, $\kappa$ can be significantly reduced. 
This is because a chain conformation in a melt is a random walk \cite{kgmodel,doibook}, 
forming several bends along the chain contour. Here, each bend acts as a scattering center for the 
heat flow. The higher the number of bends along the backbone, the lower the $\kappa$ \cite{chenkink}. 
It is also important to note that the typical end-to-end distance $R_{\rm ee}$ of a
chain with $N_{\ell}=50$ in a melt is $R_{\rm ee} \simeq 10 \sigma$, which is only 
about 15\% of the box length $L \simeq 66.2\sigma$. This $L$ to $R_{\rm ee}$ asymmetry can 
induce a dominant contribution of the vdW-dominant non-bonded interactions on $\kappa$. In this context, it is known that 
the non-bonded interactions can impact the heat flow in the polymeric materials. 
For example, in the case of the standard polymers, such as PMMA, PS, and PE, where vdW interactions are dominant 
$0.1\le \kappa \le 0.2~{\rm W/Km}$ \cite{shen10natnano,xie16mac}. Moreover, 
in the case of the hydrogen bond contacts $\kappa \to 0.4~{\rm W/Km}$ \cite{xie16mac,mukherji19prm}, 
such as PAM, PAA and PNIPAM. We also wish to highlight that all these vdW or the hydrogen bond-based systems 
have the very similar covalently bonded carbon-carbon backbone.

With increasing $n$, the difference in $\kappa$  becomes more prominent between the harmonic and the FENE bonds, 
see the comparison between black solid and open $\circ$ data sets in Fig. \ref{fig:kappalb} and also the 
comparison between the red solid and open $\triangle$ data sets in Fig. \ref{fig:kappalb}. 

A closer look at the data sets corresponding to the FENE bond further reveal that $\kappa$ remains almost invariant with $\ell_{\rm b}$,
see the data sets corresponding to open $\circ$ and $\triangle$ in Fig. \ref{fig:kappalb}.
This is predominantly due to the fact that the competing effects, i.e., the reduction in 
$\rho_{\rm b}$ with $\ell_{\rm b}$ that reduces $\kappa$ (see part a of Fig. \ref{fig:dens}) and the decrease 
in $v_{\rm void}$ with $\ell_{\rm b}$ that increases $\kappa$ (see part e of Fig. \ref{fig:snapvoid}), cancel each other. 
The later also induces a more homogeneous bond formation within the network.

It is also important to discuss how does $v_{\rm void}$ influence the $\kappa$ behavior. 
In these network structures, the protovoids usually act as the scattering centers for the heat flow,
where the most preferred heat propagation pathway is the bonded monomers along the circumference 
around the protovoids. Larger the fraction of $v_{\rm void}$, the larger the hindrance to the heat flow and thus 
the lower the value of $\kappa$.

In the case of the harmonic bonds, both tri- and tetra-functional networks show a $5 - 10\%$ increase in $\kappa$ with $\ell_{\rm b}$,
see the data sets corresponding to solid $\circ$ and $\triangle$ in Fig. \ref{fig:kappalb}. Here, it is worth
noting that a stiffer bond strength plays an additional contribution to the $\kappa$ behavior.
To further illustrate that the bond stiffness is the key factor for the above mentioned
increase in $\kappa$, we have performed one more set of simulations where $\epsilon/2s^2$ is reduced by a factor of 6.
Here, $\kappa/\kappa_{\rm linear} \simeq 1.81$ for all four $\ell_{\rm b}$, which is about comparable to the 
FENE bond data for $n=4$, see the open black $\circ$ data set in Fig. \ref{fig:kappalb}.

\subsection{Thermal conductivity, heat capacity and sound wave velocity}

\begin{figure}[h!]
\includegraphics[width=0.43\textwidth,angle=0]{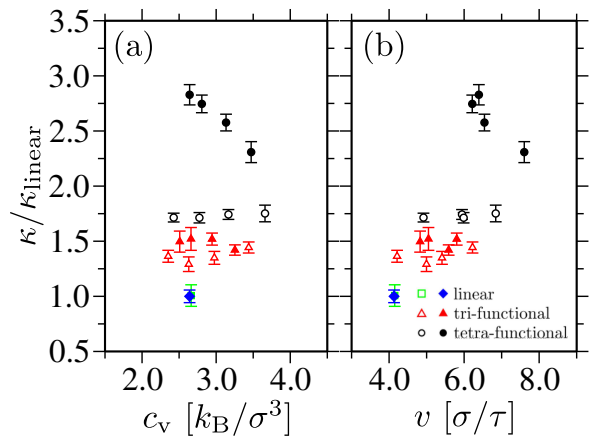}
	\caption{The normalized thermal transport coefficient $\kappa/\kappa_{\rm linear}$ as a function of 
	the volumetric specific heat $c_{\rm v}$ (part a) and the sound wave velocity $v$ (panel b).
	The data is shown for different functionality $n$, bond length $\ell_{\rm b}$ and for both bonds. 
	The data is normalized with respect to the linear chain connected by the harmonic 
	springs giving $\kappa_{\rm linear} = 5.4\pm0.3~k_{\rm B}/\tau {\rm m}$.
	Here the linear chain lengths are chosen as $N_{\ell}=50$.
	Open and solid symbols are for the FENE and the harmonic bonds, respectively.
        Lines are drawn to guide the eye.
\label{fig:kappacvv}}
\end{figure}

So far we have discussed the behavior of $\kappa$ with respect to the network microstructure and bond stiffness. 
Moreover, it has been shown that $\kappa$ is related to the velocity of sound wave $v$ and the volumetric specific heat $c_{\rm v}$ \cite{cahill92prb}.
For this purpose, we have used the simplified estimate $v = \sqrt{K/\rho_{\rm m}}$, where $K$ is
the bulk modulus. Here, $K$ is calculated using the fluctuation of volume ${\rm V}$ in the isobaric ensemble following the relation,
\begin{equation}
	K = k_{\rm B}T \frac {\left<{\rm V}\right>}{\left<{\rm V^2}\right>-\left<{\rm V}\right>^2}.
\end{equation}
The volume fluctuation is sampled over a time $5 \times 10^4 \tau$ with a time output interval of $0.5\tau$.
$c_{\rm v}$ is estimated using the Dulong-Petit limit $3\rho_{\rm m}k_{\rm B}$.
In Fig. \ref{fig:kappacvv} we show the variation in $\kappa$ with $v$ and $c_{\rm v}$.
It can be seen that these data sets show rather non-trivial functional dependence. This behavior is also
consistent with the experimental data on HCP \cite{cahill20acs}. Furthermore, this behavior
is predominantly dictated by the delicate balance between the competing effects of 
$\rho_{\rm b}$, $\rho_{\rm m}$, $v_{\rm void}$ and bond stiffness on $\kappa$. For example,
in the case of $n=4$ (or the tetra-functional HCP) and the harmonic bonds, $\kappa$ increases with
decreasing $v$ and $c_{\rm v}$, which is an effect of decreasing density and homogeneous 
bond formation in the networks, see black solid $\circ$ data set in Fig. \ref{fig:kappacvv}.

\subsection{Possible comparison with the realistic systems}

The generic simulation data presented in this study highlight the importance of bond engineering and network microstructure on
the heat management in HCP. However, the major question still remains if the results presented here can be compared 
with the experimentally relevant realistic systems. 
In this context, it is important to mention that the increased bonding in HCP ideally provides
a suitable pathway for the faster heat flow. This would, therefore, mean an increase in $\kappa$ by a large 
fraction compared to the linear polymeric materials. Here, however, we only find an increase in $\kappa$ of about a factor of 
1.5 for the tri-functional and 1.7$-$2.7 for the tetra-functional HCP, see Fig. \ref{fig:kappalb}. 
Now considering the standard linear polymeric materials, such as PMMA and PS, 
$\kappa \simeq 0.1 - 0.2$~W/Km \cite{shen10natnano,xie16mac,mukherji19prm} and the experimental data for HCP 
in Ref.~[\onlinecite{cahill20acs}], we find that $\kappa/\kappa_{\rm linear} \sim 1.1 - 2.0$ for different cross-linkers. 
This range is consistent with all the data sets presented in Fig. \ref{fig:kappalb} 
except for the tetra-functional HCP with the harmonic bonds, see the black solid $\circ$ data 
set in Fig. \ref{fig:kappalb}. Furthermore, most common HCP are 
synthesized either with fluffy bonds and monomers \cite{palmese14jmat} and/or with fluffy monomers and stiff 
cross-linkers \cite{cahill20acs}. In our model, the FENE bond type mimics these conditions closely and thus
shows reasonably good agreement with the experimental observations.

In some cases, cross-linking can also decrease $\kappa$ compared to the linear chains,
such as the cross-linked PAA system compared to the linear PAA \cite{xie16mac}.
Here, however, the length of the cross-linkers are rather large and thus 
can have large entropic fluctuation. Recent all-atom simulation results have shown that the length of 
the cross-linkers significantly contribute to the observed trends in $\kappa$ behavior.
For example, when a long PAA chain system is blended with PAM trimers that can for hydrogen-bonded 
cross-linking between two PAA monomers of the different chains, $\kappa$ can slight enhance in comparison to a 
pure PAA \cite{mukherji19mac}. These experimental and all-atom simulation data are also 
consistent with the generic simulations \cite{huo20pccp}.

Another possible route for the synthesis of HCP might be to use the linear polymer chains, 
either homopolymer or copolymer, and cross-linking them using the stiff N,N'-methylenebis(acrylamide) (BIS) \cite{backes17acs}.
In this system, it has been shown that a cross-linked microgel of P(NIPAM-co-AA) with only about 5\% BIS can 
significantly increase the elastic modulus \cite{backes17lang}. It would, therefore, be interesting to experimentally 
investigate a system with a large degree of BIS cross-linking that may give a further increase in $\kappa$ 
for network structures.

Lastly we would also like to comment on the tunability in $\kappa$. For example,
while it is always desirable to increase $\kappa$ of materials for their possible use under 
the high temperature conditions \cite{pipe14nm,cola16aami}, cross-linked thermoelectric materials \cite{liu19mme,park20apmi} 
require an ultra low $\kappa$ for better device performance. 
Here, the engineered protovoids within the cured network may provide an additional pathway for the tunability 
in $\kappa$. In this context, as discussed in the Subsection \ref{sec:netcur}, the generic features of the network microstructure 
naturally emerge because of the thermodynamic reasons.
The engineered protovoids with different fractions can be incorporated within a sample by 
including a non-reactive solvent (such as tetrahydrofuran or dichloromethane) during the network curing stage.
Inclusion of a non-reactive solvent facilitates the bond formation around a solvent bubble within 
the solution \cite{palmese14jmat}. Here, the change in non-reactive solvent content is then expected to 
change the protovoids fractions and thus $\kappa$.\\

\section{Conclusions}
\label{sec:conc}

We have performed large scale molecular dynamics simulations of two different generic polymer models
to study the thermal transport in the highly cross-linked polymers (HCP). 
We emphasize on the importance of engineered cross-linking bond types that in turns give an additional pathway to 
introduce a tunability in the thermal transport coefficient $\kappa$ of HCP. These results show that the 
spontaneous formation of molecular scale voids/protovoids during the curing procedure, the bond density, 
the bond length and the bond stiffness, together with the delicate balance between these competing 
effects ultimately dictate the behavior of $\kappa$. While our simulation data sets present a generic 
physical picture and the importance of the underlying network microstructure, we also present a 
comparative discussion in the context of the different experimentally relevant systems. 
Based on our analysis we also sketch a set of directions that may be useful for the design 
of future materials with advanced functional applications.\\

\noindent{\bf Acknowledgement:}
D.M. thanks the Canada First Research Excellence Fund (CFREF) for the financial support.
Simulations are performed at the ARC Sockeye facility of the University of British Columbia, 
the Compute Canada facility and the Quantum Matter Institute LISA cluster. 
M.K.S. thanks IIT Kanpur initiation grant for providing financial support and the computational facilities 
to create and test the input scripts and initial trajectories for simulations.

\end{document}